\documentclass[epsfig,12pt]{article}
\usepackage{graphicx}
\begin{document}
\begin{center}
\Large {The neutron 'thunder' accompanying the extensive air shower}\\
\end{center}
\vspace {0.5cm}  
\begin{center}
\footnote{E-mail address: erlykin@sci.lebedev.ru} 
A.D.Erlykin
\end{center}
\begin{flushleft}
{\em P. N. Lebedev Physical Institute, Moscow, Russia}\\
\end{flushleft}

\begin{abstract}
Simulations show that neutrons are the most abundant component among extensive air 
shower hadrons. However, multiple neutrons which appear with long delays in neutron 
monitors nearby the EAS core (~'neutron thunder'~) are mostly {\em not} the neutrons of
 the shower, but have a secondary origin. The bulk of them is produced by high energy 
EAS hadrons hitting the monitors. The delays are due to the termalization and diffusion
 of neutrons in the moderator and reflector of the monitor accompanied by the 
production of secondary gamma-quanta. This conclusion raises the 
important problem of the interaction of EAS with the ground, the stuff of the detectors
 and their environment since they have often hydrogen containing materials like 
polyethilene in neutron monitors. Such interaction can give an additional contribution 
to the signal in the EAS detectors.   
It can be particularly important for the signals from scintillator or water tank 
detectors at {\em km}-long distances from the EAS core where neutrons of the shower 
become the dominant component after a few $\mu s$ behind the EAS front.
\vspace{1pt}
\end{abstract}

\section{Introduction}
The dispute on the role of low energy neutrons as the possible origin of delayed 
(~'sub-luminal'~) pulses in neutron counters and scintillator extensive air shower
(~EAS~) detectors started long ago \cite{Tong1,Tong2,Greis,Linsl}. In those early works
 people observed single pulses delayed with respect to the main shower front and 
concluded that they are produced by neutrons accompanying EAS without specifying their 
origin in more detail. The observed delays in these detectors were not dramatic, 
however, and did not exceed a few $\mu s$. 

The present work has been inspired by the 
observation of the {\em multiple} neutrons which followed EAS with 
 delays as long as {\em hundreds} of $\mu s$ \cite{Chub1,Aushe,Anto1,Anto2,Chub2}. 
Such delays have been observed with the Tien-Shan neutron monitor for EAS in the PeV
energy region. Later this finding has been confirmed by other experiments 
\cite{Sten1,Sten2,Gawin,Baygu,Jedrz} and 
the existence of the effect is now beyond any doubts. There is, however, no 
agreement about its origin \cite{Jedrz,Sten3,Chub3}. Briefly the essence of 
the effect is the appearence of the numerous neutrons delayed by hundreds of 
microseconds after the passage of the main shower disk in the vicinity of the EAS core. 
In the spectacular scenario of 'the thunderstorm model' this phenomenon has been 
compared with the thunder which appears with a delay after a strike of the lightning
 during the thunderstorm \cite{Ambro}. 

In the detailed study 
\cite{Anto2} of the effect it has been claimed that the process has a threshold and 
the delayed neutrons appear at PeV primary energies, i.e. in the region of the 'knee' 
observed in the primary cosmic-ray (~CR~) energy spectrum. Another observation is 
that these neutrons are concentrated within a few meters around an EAS core and 
accompanied by delayed $\gamma$-quanta. The neutron multiplicity spectrum changes its 
slope at high energies and becomes flatter than at low energies. All these
features let the authors assume that this phenomenon is connected with the properties 
of high energy (~PeV~) interactions and it 'distinctly conflicts with the modern EAS 
development models of a quasi-scaling type' \cite{Chub3}. 

The other groups \cite{Jedrz,Sten3} argued that the delayed neutrons appeared not in 
the process of EAS development in air. They attributed the effect mostly to the low 
energy physics and explained it by neutrons which are produced inside the neutron 
monitor by numerous nuclear scatterings and disintegrations, caused by hadrons in the 
EAS core, and which then propagate outside the core region. Some of them appear also as
 albedo neutrons from the nuclear cascade developing in the ground underlying the 
neutron detectors after EAS propagate from the air into the ground. That explanation 
has been based on their own experimental data. 

Below we present our viewpoint on the origin of observed effects.
We shall also discuss possible consequences of our explanation for the experimental
observation of EAS at higher energies, viz. in the EeV energy range. 

\section{The PeV energy range}

\subsection{The change of the temporal distribution} 

In our paper \cite{Ambro} we showed that the pure geometrical approach based on an 
assumption that all neutrons are produced in the EAS core and propagate outside by the 
spherical diffusion cannot give an explanation of the distorted temporal
distributions at high energies and the longitude of the delay.

The distributions presented in \cite{Anto2} indicate that their distortions have most
likely a methodical explanation connected with the 'saturation' of neutron counting 
rate at the high neutron intensities. The arguments for that conclusion are the 
following: \\ 
(i) the maximum 'saturation' counting rate of the standard neutron monitor unit with 6 
counters of SNM15 type (~the dead time $\tau \approx 2 \mu s$~) is expected as 
$6/\tau \approx 3 \mu s^{-1}$; \\
(ii) this saturation is expected to start close to the observed neutron multiplicity
$M = 3 \mu s^{-1}/f(0)$, where 
$f(t) = \frac{0.72}{\tau_1}exp(-t/\tau_1) + \frac{0.28}{\tau_2}exp(-t/\tau_2)$ is the
standard temporal distribution function for neutrons produced instantly inside the 
neutron monitor and detected by its counters. For $\tau_1 = 250 - 300 \mu s$ and
$\tau_2 = 600 - 650 \mu s$ \cite{Aushe} the saturation level at $3\mu s^{-1}$ can
be reached at $logM = 2.9 - 3.0$. 

These expected values are exactly what is observed in the experiment \cite{Anto2}.
Both arguments as well as the experimental study of the phenomenon made with neutron 
detectors which have better time resolution \cite{Sten1,Sten2,Sten3} give strong 
support to the methodical explanation of the observed distortions of the time 
distributions which are caused by the high neutron intensity inside the monitor and the
 saturation of the counting rate due to the finite time resolution of neutron counters.

The appearance of the minimum in the neutron temporal distributions at delays of 200 - 
300 $\mu s$ at high neutron intensity seen in SNM18 counters and not seen in 
SNM15 counters evidences also in favor of the internal, probably methodical origin of 
this effect, not connected with EAS. The similar explanation is proposed by authors 
themselves \cite{Chub2}.

\subsection{The concentration of the effect in the region near the EAS core}

If the distortions of the neutron temporal distribution are due to the saturation of 
the counting rate this effect should be observed in the region with the highest density
 of neutrons, i.e. near the EAS core. Indeed distortions are observed only in one 
module of the neutron monitor closest to the EAS core. Remarkable is a very high 
counting rate in this module. The distortion starts to be seen at the recorded 
multiplicity ( in 0 - 3400 ns interval ) of $M\approx$ 630-1000. The measured efficiency of
 the monitor is 5-6\% \cite{Anto2}, so that the total number of neutrons produced in 
the module is $N \approx (1-2)\cdot 10^4$. This number is produced by EAS of an 
approximately PeV energy. For showers in which the counting rate achieves the 
saturation level at the end of the measurement interval, i.e. at 3400 ns, the produced 
number of neutrons can be higher by the factor of 2000 (!). Similar estimates have been
 made in \cite{Chub2}. 

It will be shown later that the spatial distribution of neutrons associated with a 
shower is rather wide. The high concentration of the observed neutron density around 
the EAS core indicates that these neutrons are mostly of secondary origin, i.e. 
produced inside the neutron monitor and do not arrive from outside.

In order to understand the phenomenon we have made calculations. In contrast with 
\cite{Ambro} where we used an analytical approach, we have made here more adequate 
Monte Carlo simulations.      
Even with the Monte Carlo approach it is complicated to reproduce precisely the 
conditions of the experiment \cite{Anto2}. Just to get an estimate of the phenomenon we
 simulated showers from 1 PeV primary protons with the vertical incidence upon the 
atmosphere observed at the Tien-Shan altitude of 3340 
m a.s.l. (~687 gcm$^{-2}$~). We used CORSIKA6500 code \cite{Heck} with QGSJET-II high 
energy interaction model \cite{Ostap} and Gheisha 2002d for low 
energy interactions below 80 GeV \cite{Fesef}. NKG version has been used for the 
electromagnetic component, energy thresholds were taken as minimum recommended by 
\cite{Heck}, i.e. 0.05, 0.05, 0.001 and 0.001 GeV for hadrons, muons, electrons and 
photons respectively. We determined characteristics of particles both for the total 
shower and for those which hit the monitor of $3\times 2m^2$ area at its center.  

The lateral distribution of neutrons compared with that of main other hadron components
 - protons and pions, is shown in Figure 1a and the relevant total number of particles 
- in Table 1. The energy spectrum of neutrons in the whole shower and those which hit 
the monitor is shown in Figure 1b. I underline again that these neutrons and hadrons 
are {\underline{shower}} neutrons and hadrons, which are produced in the air, but 
{\em not} in the monitor or in the ground.  
\begin{figure}[htbp]
\begin{center}
\includegraphics[width=7.4cm,height=7.4cm,angle=0]{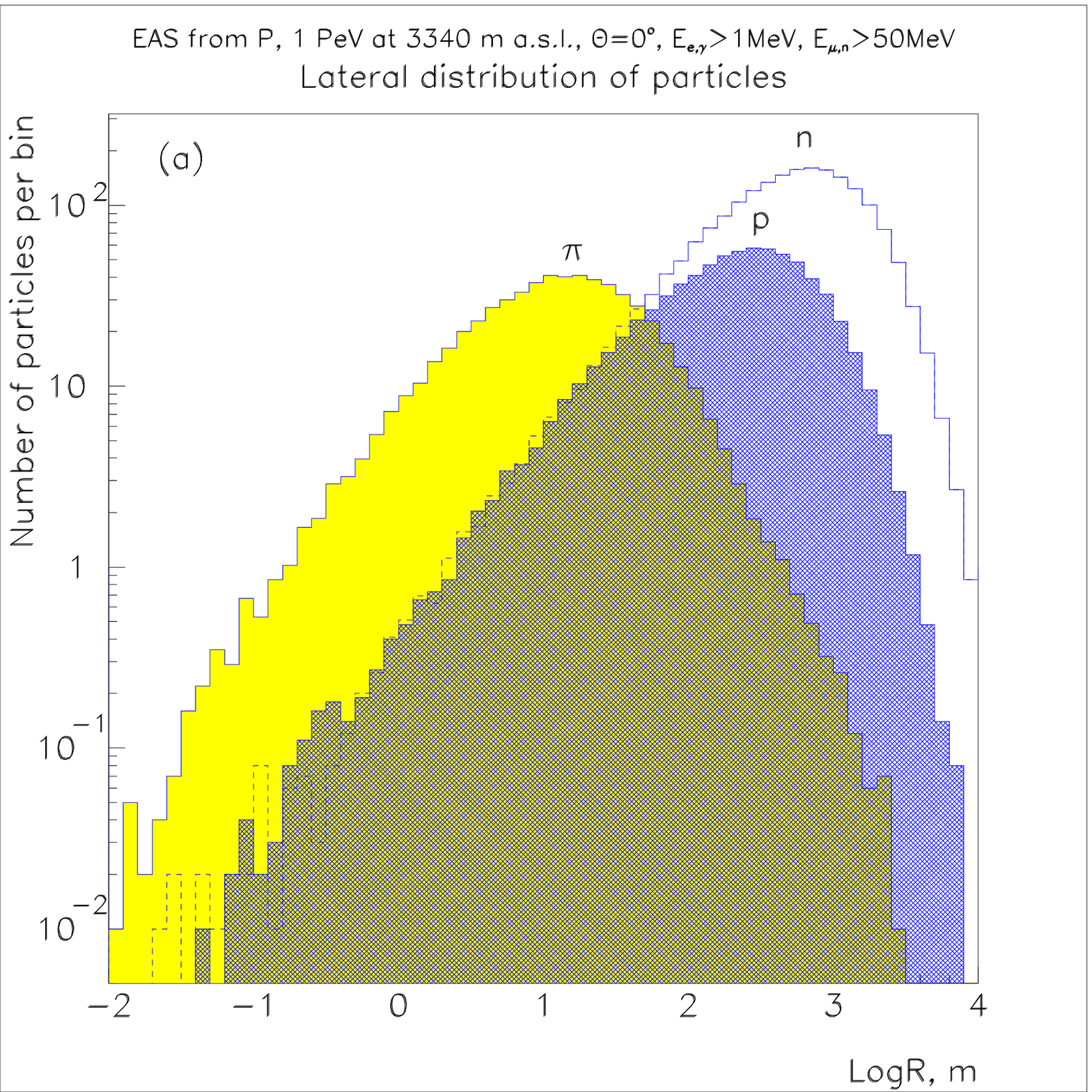}
\includegraphics[width=7.4cm,height=7.4cm,angle=0]{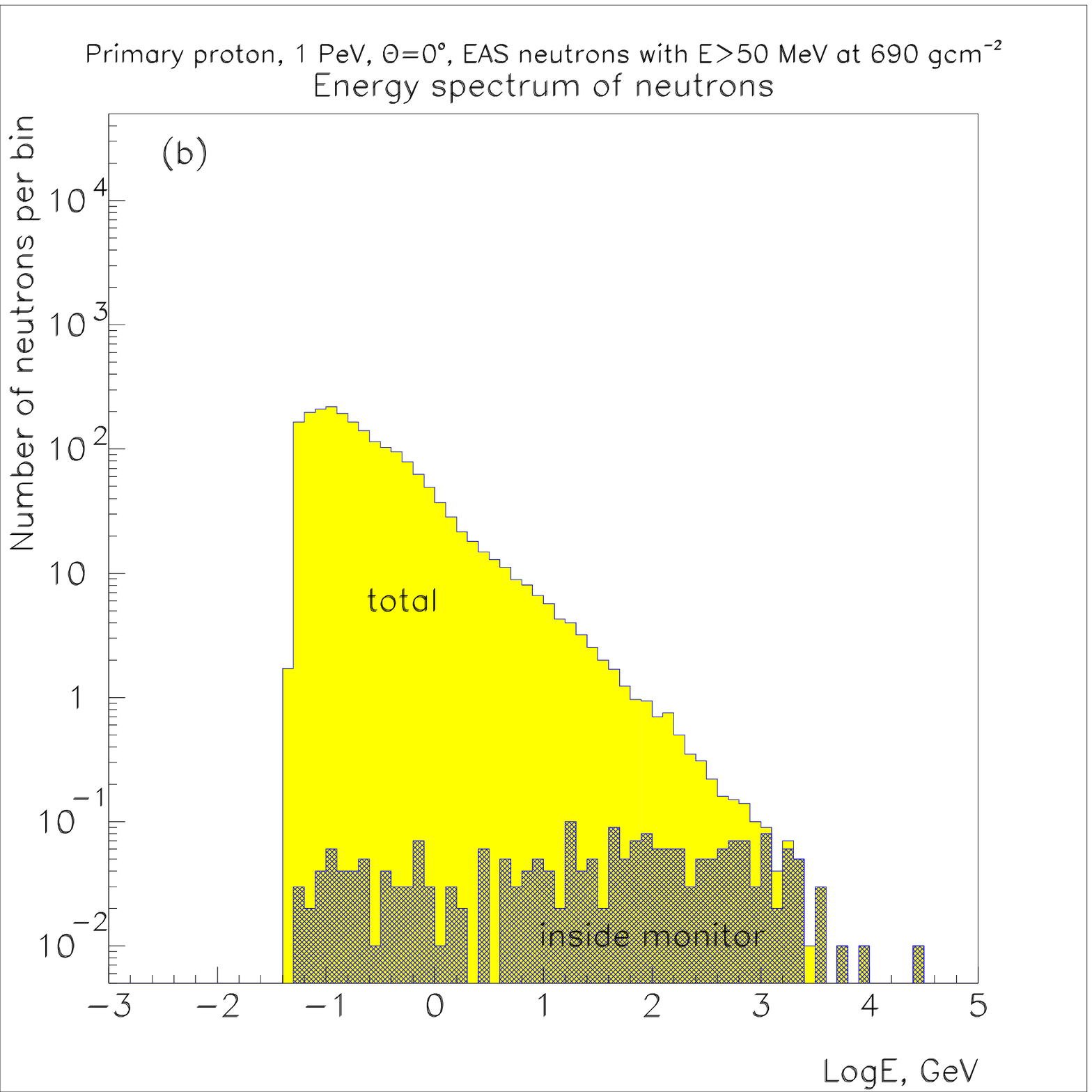}
\caption{\footnotesize Characteristics of hadrons in EAS from 1 PeV primary proton 
incident vertically on the level of 3340 m a.s.l.: (a) Lateral distribution of protons,
 neutrons and pions with energies above 50 MeV; (b) Energy spectrum of neutrons in the 
total shower and of those hitting the neutron monitor of $3\times 2m^2$ area.}
\end{center}
\label{fig:neu1}
\end{figure}

Remarkable is that the neutron component is the most abundant among all nuclear-active 
particles. The total number of neutrons above 50 MeV in such showers is about 2000,
while protons amount as much as ~750 and pions ~600. The energy spectrum presented in 
Figure 1b, shows that the bulk of neutrons are low energy neutrons. Their differential 
energy spectrum is of $~E^{-2}$ type, but it continues up to a few tens of TeV (!) 
This spectrum shape gives a possibility that if the energy threshold of 50 MeV could be
 reduced to, say, 1 MeV, the total number of neutrons will be substantially higher. 
However these low energy neutrons populate mostly the periphery of the shower at 
distances up to kilometers from the axis. As it is seen from the Table 1 in the EAS 
core $\sim$1.6\% of all hadrons including neutrons carry $\sim$50\% of the whole energy
 of hadrons in the shower. Since most of neutrons recorded by the monitor have the 
thermal energy of 0.02 eV even such a huge total number of them as $(1-2)\cdot 10^4$ 
carry the total energy no more than 0.2-0.4 KeV, which is incomparably small with 
$\sim$40 TeV provided by EAS hadrons in the core and should not create 'the energy 
crisis' during their cascading with the subsequent production and thermalization of 
neutrons. Simulations of neutrons in EAS have been also made in \cite{Bouri}.   

As it can be seen from Table 1, if such a shower hits the neutron monitor at the centre
 of one of its module with $3\times 2m^2$ area the mean number of hadrons inside this 
module is about 55 and their total energy is about 40 TeV. The lead in the monitor 
is a neutron rich element and it is poor absorber of fast neutrons. Interactions of 
hadrons with lead are followed by emission of recoil and evaporation neutrons with  
their subsequent moderation and thermalization inside the polyethilene reflector. This
process leads to an accumulation of multiple neutrons inside the monitor. If to take 
the results of calibration for this neutron monitor connecting the hadron energy 
$E_h$ and the multiplicity of {\em produced} neutrons $N$ as $N = 35\cdot E_h^{0.5}$ 
with $E_h$ in GeV \cite{Aushe}, then 40 TeV of hadron energy, which hit the monitor, 
give  $0.7\cdot 10^4$ neutrons. This estimate is in good agreement with an 
estimate for the number of neutrons $\sim (1-2)\cdot 10^4$ created inside the monitor 
by EAS of PeV energy for which the saturation of the counting rate has just started to 
be seen.  
\begin{center}
\begin{tabular}{||c||c|c|c||}                                                             \hline\hline
\bf{       } &{\em total in EAS} & {\em monitor at R=0} & {\em monitor at R=4m}   \\ \hline\hline
   {$ N_h$}      &  3445         &        55            &     9.3                 \\ \hline
   {$E_h$, GeV}  & 72610         &     37680            &   823.5                 \\ \hline
   {$N_n$}       &  1994         &         2.1          &     0.6                 \\ \hline
   {$E_n$, GeV}  &  3389         &      1115            &    25.1                 \\ \hline\hline
\end{tabular}
\end{center}
{\footnotesize Table 1. The number and the energy content of hadrons with the energy 
above 50 MeV in the EAS from the 1 PeV primary proton incident vertically on the 
observation level of 3340 m a.s.l.. $N_h$, $E_h$ and $N_n$, $E_n$ are the number and 
the energy in GeV for all hadrons and neutrons respectively for the whole shower and 
for those hitting the neutron monitor of $3\times 2m^2$ area located at the EAS axis or
 4m apart.}

Table 1 demonstrates also the strong concentration of energy in the core region:
if the monitor is at 4m from the shower axis the energy of hadrons and neutrons 
hitting it decreases by the factor of 40. 
   
    These estimates show that the numerous neutrons
which are detected in the neutron monitor after the passage of EAS are {\em not} the 
EAS neutrons which are produced in the air and follow the main shower disk with delays 
no longer than a few $\mu s$, but neutrons {\em produced in the lead} inside the 
monitor during the process of inelastic scattering of high energy EAS hadrons with 
their subsequent attenuation and thermalization within a polyethilene moderator and 
reflector which is a relatively long process lasting hundreds of $\mu s$.    

\subsection{Delayed $\gamma$-quanta}

 As for the existence of gamma-quanta delayed by 
hundreds $\mu s$ their intrinsic connection and an origin from neutron induced 
reactions seem to be apparent. The authors of \cite{Anto2,Chub3} notice themselves 
the similarity of gamma-quanta temporal distribution with that of neutrons. The rapid 
rise of their integral number with the neutron multiplicity $M$ at $M > 630$, mentioned
 in \cite{Anto2}, is most likely {\em not} due to the real fast rise, but on the 
opposite - with a slower rise of recorded $M$ compared with the rise of the true 
corrected multiplicity $N$ due to the saturation effect mentioned above. The small 
counting rate at the beginning of the measurement time interval and its maximum at 
about hundred $\mu s$, observed at large $M$ is most likely connected with the long 
recovery time of gas counters after a passage of high fluxes of particles in large EAS.

Therefore the gamma quanta delayed by hundreds of $\mu s$ do not evidence for the 
new processes in the development of EAS in PeV energy region. The bulk of them have a 
local origin from neutrons created inside the neutron monitor. 

\subsection{The flat spectrum of the 'true' neutron multiplicity}

In \cite{Chub3} it is argued that the spectrum of the 'true' neutron multiplicity $N$
derived from the observed multiplicity $M$ by correcting it for the saturation effect 
is too flat to be explained by this methodical reason. Fitted by the power law it has 
the shape of $dI/dN \sim N^{-2}$ above $M \approx 1000$, which contradicts to the 
expected shape $\sim N^{-3}$ at energies above the knee. I think that on the contrary 
this observation supports the assumption that the distortion of the temporal 
distribution is due to the saturation of the counting rate. 

Indeed in the case of a saturation the observed and the 'true' neutron multiplicity are
connected as \cite{Golda}
\begin{equation}
M = \frac{N}{1+N\frac{\tau}{T}}
\end{equation}
where $\tau$ is the dead time and $T$ is the collection time during which neutrons are
recorded.
The differentiation of the equation (1) gives $dM = \frac{dN}{(1+N\frac{\tau}{T})^2}$ 
and if $dI/dM \sim M^{-\gamma}$ then
\begin{equation}
\frac{dI}{dN}(N)=\frac{dI}{dM}(N) \frac{dM}{dN}(N)=(\frac{N}{1+N\frac{\tau}{T}})^{-\gamma}(1+N\frac{\tau}{T})^{-2} = N^{-\gamma}(1+N\frac{\tau}{T})^{\gamma-2}
\end{equation}   
At large $N$ when $N\frac{\tau}{T} >> 1$ $dI/dN \sim N^{-2}$ for any slope of
$\gamma$. 

Incidentally $N$ exceeds $M$ by the factor of 2 at $M \sim 600-1000$ which confirms 
the validity of the expression (1) for $\tau = 2 \mu s$ and $T = 3400 \mu s$, which connects
$M$ and $N$ with an account for the saturation effect.
 
\subsection{Pre-conclusion I}
 
The presented analysis shows that our interpretation of the 
phenomenon of 'neutron thunder', based on the analysis of the experiment \cite{Anto2}
and on our own simulations and quantitative estimates is identical to the 
interpretation \cite{Jedrz,Sten3} based 
on their own experimental data. The neutrons which are observed in EAS with hundreds 
of $\mu s$ delay after the main shower front are not the result of a new physics, 
indicating the production of a new EAS component in the air and the change in the EAS 
spatial and temporal development, but is a local phenomenon, which appears as a result 
of the EAS interaction with the stuff of detectors and their environment.\\ 
1. The distortion of the temporal distribution is caused by the saturation of the 
counting rate at high neutron fluxes due to the finite time resolution of counters.\\
2. The observed threshold of the effect is right at the energy when this saturation
is achieved.\\
3. The concentration of the effect in the region of the EAS core is due to the steep 
lateral distribution of EAS hadrons and their energy.\\
4. The delayed gamma-quanta have the secondary origin produced by delayed neutrons.\\
5. The flat spectrum of 'true' neutron multiplicities is also explained by the 
saturation of the counting rate.
   
This interpretation does not understate the importance of the real discovery of the 
effect made by Chubenko A.P. with 
his colleagues. If the presented interpretation is true, the scenario of 'the neutron 
thunder' complements our knowledge of the EAS development and its interaction with the 
ground, surface detectors and their environment. Once again like in the case of the 
transition effect, when some 
part of invisible gamma-quanta is converted into electrons or electron-positron pairs 
in the thick scintillators or in thick water cherenkov detectors this effect indicates 
that our records depend on our detectors. Within this scenario another problem appears 
- the production and propagation of neutrons created when the EAS core hits the ground.
 Due to their long propagation length \cite{Agafo} these neutrons  can give observable 
effects both at shallow depths underground in 
particular at mountain altitudes where the EAS cores are more energetic, and as albedo 
neutrons - in surface detectors. There might be plenty of other interesting effects 
worth of the experimental and theoretical study. 

\section{The EeV energy range}
 
\subsection{The role of neutrons at large EAS core distances}

 In the light of the possible 
contribution of neutrons to the signal in water and ice cherenkov 
detectors used in many experiments, viz. Pierre Auger Observatory (~PAO~), MILAGRO, 
NEVOD, Ice-Top and others, we have made simulations of EAS in EeV energy range similar 
to those made at PeV energies. The interaction model is the same QGSJET-II, but the 
observation level is that of PAO, i.e. 1400 m a.s.l.. The primary proton has 1 EeV 
energy, the electromagnetic component has been simulated using EGS4 option with the 
thinning level of $10^{-5}$. 

The lateral distribution of electromagnetic component (~$e^+ + e^- + \gamma$~), muons 
 (~$\mu^+ + \mu^-$~) and neutrons is shown in Figure 2. The gamma-quanta are included 
into the graph since the water tanks of PAO are thick enough to absorb them and to get 
their contribution to the 
signal. Both distributions of particles number and of particles energy are shown. 
The separation of PAO water tanks is 1.5 km. It is seen that {\em if} neutrons are 
absorbed in the water and their energy is transformed into the visible light they could
 contribute up to 10\% to the signal at such large distances.       

Correlation plots of different lateral, angular, temporal and energy characteristics of
 EAS neutrons are shown in Figure 3.
\begin{figure}[htb]
\begin{center}
\includegraphics[width=8cm,height=8cm,angle=0]{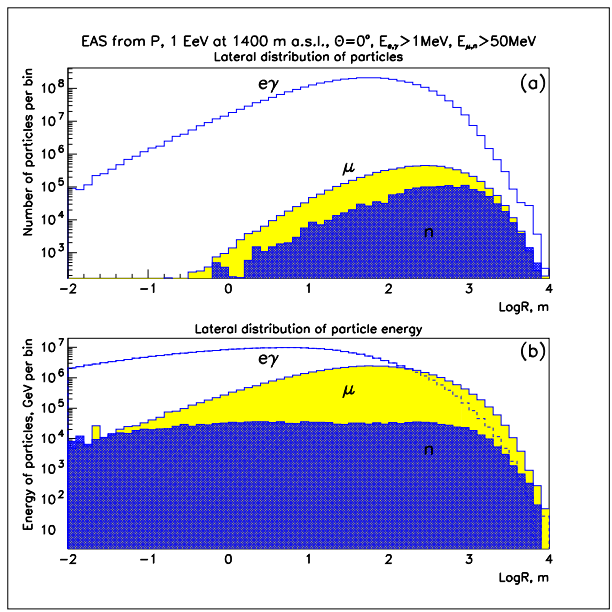}
\caption{\footnotesize Lateral distribution of particle numbers (a) and the particles 
energy
(b) for 1 EeV primary proton incident vertically at the level of 1400 m a.s.l. It is 
seen that neutrons could contribute up to 10\% to the signal of water tanks at 1km 
distance from the shower axis, if among products of their interaction with water are 
relativistic electrons.}
\end{center}
\label{fig:neu2}
\end{figure}

\begin{figure}[htb]
\begin{center}
\includegraphics[width=8cm,height=8cm,angle=0]{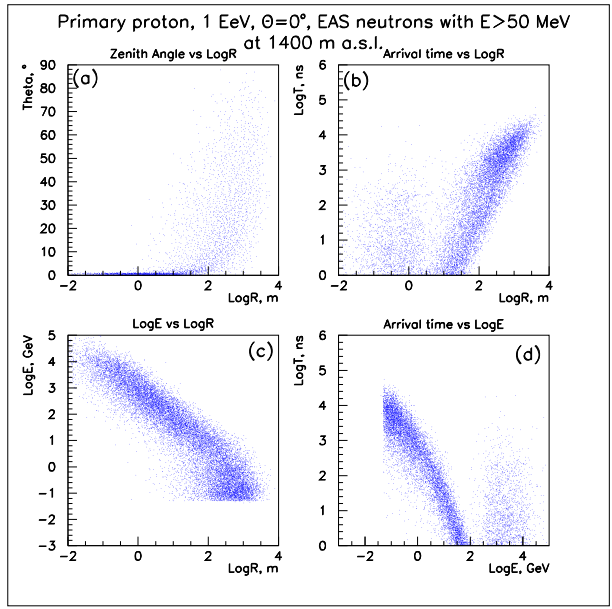}
\caption{\footnotesize Correlation plots between different characteristics of EAS 
neutrons: (a) zenith angle $\theta$ vs. core distance $R$; (b) arrival time delay $T$ 
vs. core distance $R$; (c) energy of neutrons $E$ vs. core distance $R$ and (d) 
arrival time delay $T$ vs. energy of neutrons $E$. Interesting is the distinction of 
two groups of neutrons by their energy seen in the last plot. }
\end{center}
\label{fig:neu3}
\end{figure}

It should be remarked that neutrons at large distances from the EAS core have mostly an
 energy below a few GeV and a very wide, nearly isotropic angular distribution. Their 
delays with respect to shower front spread up to tens of $\mu s$. Interestingly, 
neutrons create two distinct groups with energies above and below $\sim 10^2$GeV, seen 
clearly in Figure 3d. Apparently such separation is the consequence of different 
production mechanisms: neutrons above $10^2$GeV are produced as secondaries in high 
energy hadron collisions, lower energy neutrons appear mostly in knock-on processes.   
It is neutrons of the first group which together with other hadrons carry the bulk of 
hadron energy in the EAS core, create $\gamma$- and hadron families in X-ray films and 
ensure the subsequent multiplication process in the neutron monitor. The neutrons of 
the second group diffuse with non-relativistic speed and a wide angular distribution to
 the periphery of the shower where they can give delayed 'sub-luminal' pulses in the 
scintillators.

The temporal distribution of the particle number for electromagmetic, muon and neutron 
component at distances R$<$10m, 100 and 1000 m from the core are shown in Figure 4.
It is seen that after 5$\mu s$ at the core distance of 1km neutrons are the dominant 
component of the shower. If they could produce relativistic electrons in the process
of moderation and thermalization in water (~viz. an excitation of oxygen nuclei with 
the subsequent emission of $\gamma$-quanta or $n + p \rightarrow d + \gamma$ reaction~)
they could contribute to the signal and should be taken into account in the conversion
of S(1000) - the characteristics used by PAO for the energy estimate, into the primary
energy. The experiments on the sensitivity of water detectors to neutrons are now 
discussed.     
\begin{figure}[htb]
\begin{center}
\includegraphics[width=8cm,height=8cm,angle=0]{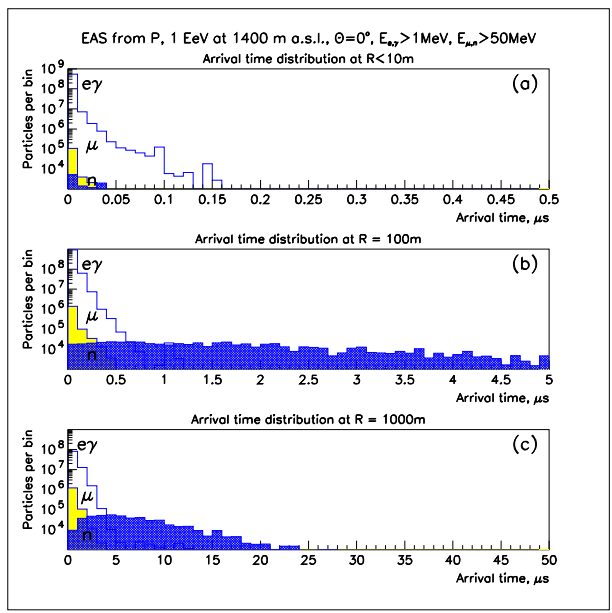}
\caption{\footnotesize Arrival time distribution of electromagnetic, muon and neutron
component of the shower at core distances less than 10m (a), 100m (b) and 1000m (c).
It is seen that at 1000m from the core neutrons dominate among other particles after
5$\mu s$.}
\end{center}
\label{fig:neu4}
\end{figure}

\subsection{Pre-conclusion II}
 
Simulations show that low energy EAS neutrons propagate far from the shower axis
and they constitute the component which has relatively long delays of their arrival 
time. At core distances approaching 1km and at about 5$\mu s$ after the passage of the 
main shower front these neutrons dominate among other particles. These distances and 
times are typical for the detectors of Pierre Auger Observatory and principally 
neutrons could contribute up to 10\% to the water tank signals. However, neutrons are 
neutral particles and at these distances they are non-relativistic, therefore they 
cannot emit cherenkov light directly. Only if in the process of their moderation and 
thermalization in water they create relativistic electrons and $\gamma$-quanta, they 
can be detected. The experiment at Tien-Shan showed that neutrons create such gamma-
quanta in the surroundings of the neutron monitor \cite{Anto2}. As for water tanks such
a possibility has to be checked experimentally.

\section{Conclusion}

Since the EAS discovery in thirties all the subsequent studies manifested the concept 
that the EAS is a complex multicomponent phenomenon. So far different detectors observe
 and study 
different EAS components: Geiger counters - charged particles, mostly electrons, thick 
scintillators are sensitive also to some part of gamma-quanta, gamma-telescopes - 
cherenkov light, ionization calorimeters - an electromagnetic and hadron component and 
X-rays - highest energy part of these components. Neutrons are neutral particles and so
 far they were not studied separately from all other hadrons. It is a merit of Chubenko
 A.P. and his colleagues who applied neutron monitors for the detailed study of the 
neutron component of EAS. The neutron monitor is the detector which includes the 
moderator and the reflector - the hydrogen containing materials, which increase the 
sensitivity of the device to neutrons. Chubenko A.P. et al. discovered the 'neutron 
thunder' - neutrons delayed up to $ms$ after the passage of the main shower front. 
Alhough according to our interpretation the bulk of the observed neutrons have a 
seconary origin, i.e. they are produced and delayed inside the monitor, the existence 
of the neutrons produced by EAS in air and accompanying the main shower front is now 
without any doubt. The true 'neutron thunder' associated only with EAS is not so long 
as that observed inside the monitor - the simulations show that it can last up to 
hundred $ns$. However neutrons of EAS can definitely cause the same effects in the
environment, in the ground and in the detectors as they make in the neutron monitor, 
like an 'echo effect', which lasts up to hundreds of $\mu s$.
 
It is particularly true for the studies at the mountain level where EAS cores are more 
energetic than at sea level and a good part of the year the ground is covered by snow  
(~Tien-Shan, Aragats, Chacaltaya, South Pole~) sometimes a few meters thick. As for the
 Tien-Shan station there might be an additional factor emphasizing the role of neutrons
 - its ground is a permafrost with a good fraction of ice inside. 

As for the detector sensitivity to neutrons, water and ice tanks are particularly worth
 of attention. First of all water is not just an absorber, but also a moderator. 
Secondly, although the neutrons as neutral and mostly non-relativistic particles cannot
 produce cherenkov light directly, the study \cite{Anto2} showed that they produce 
gamma-quanta and electrons, which can be eventually detected by water tanks due to 
their emission of cherenkov light. Since water and ice filled detectors are wide spread
 all over the world and in particular used in the Pierre Auger Observatory, the 
contribution of neutrons to their signals at large distances from the EAS core and at 
large delays from the trigger moment, can be substantial. It should be analysed and 
taken into account if necessary. The same remarks could be referred to hydrogen 
containing plastic scintillators used in many other large EAS arrays (~Yakutsk, 
Telescope Array etc.~). As has been mentioned above, signals delayed by $\mu s$ 
(~'subluminal pulses~'~) have been already observed in larhe scintillator arrays, such 
as Volcano Ranch \cite{Greis,Linsl}. 

Presumable the effect of 'the neutron thunder' can be applied in practice for 
the neutron carotage of the upper layers of the ground. Instead of the artificial 
neutron source in this method the ordinary EAS can be used since EAS cores carry on 
and produce a lot of secondary neutrons. Also 'the neutron thunder' can be used for 
the search of water on the Moon or on the surface of other planets, like it is being 
made with Mars Odyssey mission \cite{Mitr1,Mitr2}      

In any case the 
phenomenon of 'neutron thunder' complements our knowledge of the EAS development and 
it is certainly worth of further experimental and theoretical analysis. 

{\bf Acknowledgments}

The author thanks the INFN, sez. di Napoli and di Catania, personally Professors 
M.Ambrosio and A.Insolia for providing the financial support for this work and their 
hospitality. I also thank Martirosov R., Petrukhin A., Ryazhskaya O.G., Stenkin Yu.V., 
Szabelski J., Ter-Antonian S., Tsarev V.A., Vankov Kh., Watson A. and Yodh G. for useful 
discussions and references.

\end{document}